\documentclass[%
 aip,
 jmp,%
 amsmath,amssymb,
 preprint,%
]{revtex4-1}

\usepackage{graphicx}
\usepackage{dcolumn}
\usepackage{bm}

\begin{document}

\preprint{AIP/123-QED}

\title[]{Simultaneous radiation pressure induced heating and cooling of an opto-mechanical resonator}

\author{Siddharth Tallur}
 \email{sgt28@cornell.edu.}
\author{Sunil A. Bhave}%
\affiliation{ 
OxideMEMS Laboratory, Cornell University, Ithaca, NY 14853, USA
}%

\date{\today}

\begin{abstract}
Cavity opto-mechanics enabled radiation-pressure coupling between optical and mechanical modes of a micro-mechanical resonator gives rise to dynamical backaction, enabling amplification and cooling of mechanical motion. Due to a combination of large mechanical oscillations and necessary saturation of amplification, the noise floor of the opto-mechanical resonator increases, rendering it ineffective at transducing small signals, and thereby cooling another mechanical resonance of the system. Here we show amplification of one mechanical resonance in a micro-mechanical ring resonator while simultaneously cooling another mechanical resonance by exploiting two closely spaced optical whispering gallery mode cavity resonances.
\end{abstract}

\maketitle


\section{\label{sec:level1}Introduction}

Mechanical oscillators coupled to the electromagnetic mode of an opto-mechanical cavity have emerged as an important new frontier in photonics, and have enabled interesting experiments in cavity opto-mechanics. Recent work has shown resonators with mechanical displacement sensitivities close to the zero point motion of the mechanical modes \cite{kippcooling}. Optical forces have also been shown to exist in these opto-mechanical systems which can be used for a variety of applications including static motion of micro mechanical structures \cite{lipsonstatic, painterstatic}, setting up oscillations (heating) of the vibrational modes \cite{kippvahala, vahalarp, paintercrystal, carmonsbs} and cooling of vibrational modes to achieve ground state \cite{kippcooling, apselmeyer, marquadt}. The opto-mechanical systems can be either stand alone opto-mechanical resonators on chip which are interrogated by evanescent coupling from fiber taper that provide lower insertion loss \cite{vahalarp, paintercrystal, carmonsbs} or on-chip systems which incorporate waveguides along with opto-mechanical resonators on the same chip \cite{lipsonstatic, sidsinomo}.

The ultimate sensitivity of optical sensing of mechanical motion is fundamentally set by the Standard Quantum Limit (SQL) \cite{bowen}. However, well before the SQL is reached, backaction forces may dominate and severely alter the dynamics of the intrinsic mechanical motion of the sensor. Due to a combination of large mechanical oscillations and necessary saturation of amplification, the noise floor of the opto-mechanical sensor increases, rendering it ineffective at transducing small signals. Parametric instability is predicted to be a potential problem in the context of the advanced Laser Interferometer Gravitational Observatory (LIGO) \cite{ligo} and, more generally, in many cavity opto-mechanical systems designed for ultra-precise sensing. This can be controlled by designing elaborate feedback schemes \cite{ligo, bowen}. Here we show amplification of one mechanical resonance in a silicon nitride micro-mechanical ring resonator while simultaneously cooling another mechanical resonance by exploiting two closely spaced optical whispering gallery mode cavity resonances. The possibility of simultaneous heating and cooling can open up avenues in studying coherent phonon exchange and phonon dynamics between different acoustic modes, and be of interest in MEMS gyroscopes and studying aspects of condensed-matter and many-body physics at the macro-scale.

\begin{figure}[htbp]
\centering
\includegraphics[width = 8.5cm]{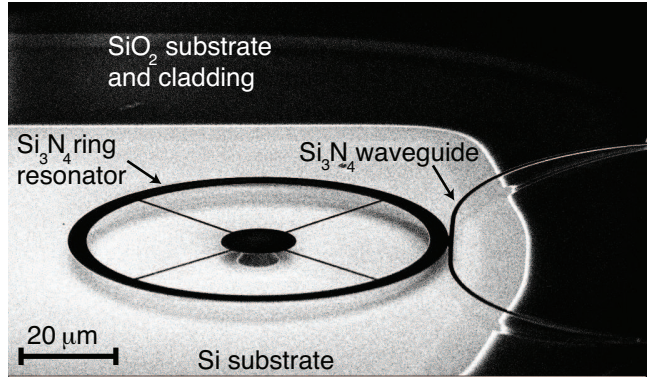}
\caption{Scanning electron micrograph (SEM) of the released opto-mechanical resonator. The silicon nitride ring thickness is 300nm and it has an outer radius of 40$\mu$m and width of 6$\mu$m. The distance between the resonator and the waveguide at the point of closest approach is 50nm. The released silicon nitride ring is supported by spokes of width 500nm and supported on a central silicon dioxide pedestal.}
\label{sem}
\end{figure}

\section{\label{sec:level1}Device fabrication and design}
The opto-mechanical ring resonator is designed in silicon nitride and supported on a silicon dioxide pedestal. The  micro-mechanical ring has an outer radius of 40$\mu$m and width of 6$\mu$m. We use grating couplers and an integrated waveguide of width 800nm to couple light evanescently to the resonator. To fabricate the device, we start with silicon wafers that have 4$\mu$m silicon dioxide thermally grown and deposit 300nm silicon nitride using low pressure chemical vapor deposition (LPCVD). The resonators, waveguides and grating couplers are defined using electron beam lithography. The pattern is transferred into the nitride device layer using CHF$_3$/O$_2$ reactive ion etch. We deposit SiO$_2$ cladding using plasma enhanced chemical vapor deposition (PECVD), to reduce grating transmission losses. A second mask is then used to pattern release windows near the resonator using contact photolithography. This is followed by a partial etch into the cladding and a timed release etch in buffered oxide etchant to undercut the devices. The samples are then dried using a critical point dryer to prevent stiction. The resulting devices have cladding over the gratings, and the tapered section of the waveguide. Figure~\ref{sem} shows a scanning electron micrograph (SEM) of the released resonator.

\section{\label{sec:level1}Opto-mechanical interactions}
\begin{figure}[htbp]
\centering
\includegraphics[width = 8.5cm]{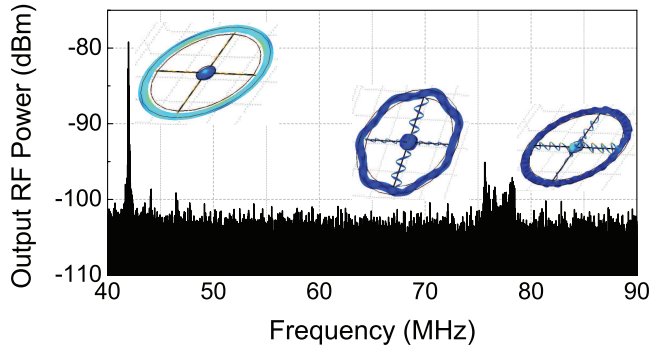}
\caption{RF spectrum at the output of the avalanche photodetector. The peaks observed correspond to the Brownian noise mechanical motion of the micro-ring. The fundamental radial expansion mode of the micro-ring at a frequency of 41.97MHz causes strong intensity modulation of the laser light as compared to a group of azimuthal composite mechanical modes around 77MHz.  An optical resonance with total optical quality factor of 200,000 was used to probe the opto-mechanical response of the resonator. The laser power used was 5dBm. The input and output grating couplers each introduce a loss of 8dB. Finite element method (FEM) simulations of the mechanical mode shapes are shown as insets.}
\label{brownian}
\end{figure}

We probe the interaction between the optical and mechanical modes of the ring resonator using an avalanche photodetector to convert motion induced intensity modulation into RF signals \cite{sidsinomo}. We choose an optical mode with an optical quality factor of $\approx$ 200,000 (and intrinsic Q $>$ 500,000) and the laser wavelength is fixed such that it corresponds to a 3dB drop in optical transmission off-resonance. At low input laser powers (5dBm), the input light coupled into the cavity is modulated by the Brownian noise motion of the mechanical modes of the micro-ring as shown in Figure~\ref{brownian}. The fundamental radial expansion mode of the micro-ring at a frequency of 41.97MHz causes strong intensity modulation of the laser light as compared to a group of azimuthal composite mechanical modes around 77MHz. This can be attributed to higher effective path length change associated with the radial expansion mode, which causes greater modulation of the laser light \cite{kippvahala}.

\subsection{\label{sec:level2}Heating and cooling of fundamental radial expansion mode}
\begin{figure*}[htbp]
\centering
\includegraphics[width = 17cm]{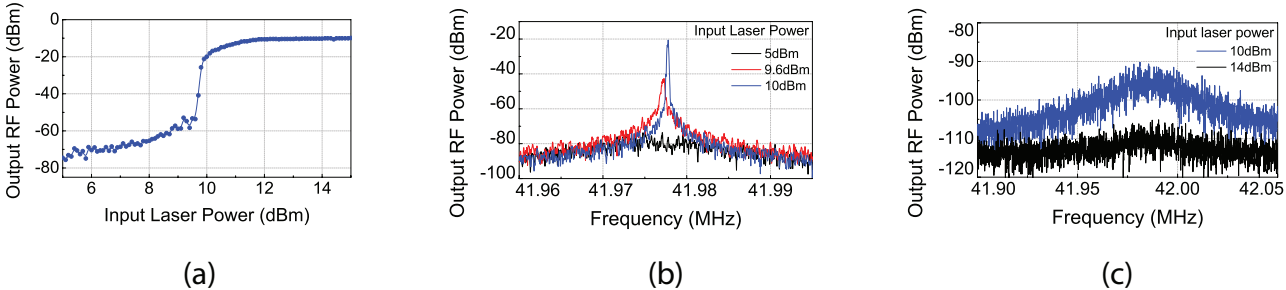}
\caption{(a) Variation of RF power at the output of the photodetector with the input laser power, for the fundamental radial expansion mode of the micro-ring at 41.97MHz. As the laser power is increased, self-sustained oscillations are observed for this mode. The sharp threshold behavior shown is characteristic of radiation pressure induced parametric instability. (b) The mechanical mode is heated by blue detuning the input laser light (1550.55nm) with respect to the cavity (1550.6nm). The linewidth of the peak narrows and the frequency increases (blue curve), as expected for heating of the mechanical mode. c) When the laser is red detuned (1550.672nm) with respect to the cavity, the mechanical mode is cooled. The linewidth increases from 42.4kHz to 92.5kHz as the laser power is increased from 10dBm (blue curve) to 14dBm (black curve). This corresponds to an effective temperature of 138K.}
\label{threshold}
\end{figure*}

The fundamental radial expansion mode of the ring at 41.97MHz has mechanical Q $\approx$ 2,000 measured in air. As we increase the laser power, self-sustained oscillations are observed for this mode above the input threshold power as shown in Figure~\ref{threshold}.a. The sharp threshold behavior is characteristic of radiation pressure induced parametric instability \cite{kippvahala}. Figures~\ref{threshold}.b and~\ref{threshold}.c show heating and cooling of this mechanical mode obtained by blue detuning and red detuning the laser with respect to the cavity respectively. The mechanical mode is heated by blue detuning the input laser light (1550.55nm) with respect to the cavity (1550.6nm). The linewidth of the peak narrows and the frequency increases, as expected for heating of the mechanical mode. When the laser is red detuned (1550.672nm) with respect to the cavity, the mechanical mode is cooled. The linewidth increases from 42.4kHz to 92.5kHz as the laser power is increased from 10dBm to 14dBm. This corresponds to an effective temperature of 138K. The effective temperature is inferred by the linewidth of the mechanical resonance \cite{temperature}.

\begin{figure*}[htbp]
\centering
\includegraphics[width = 17cm]{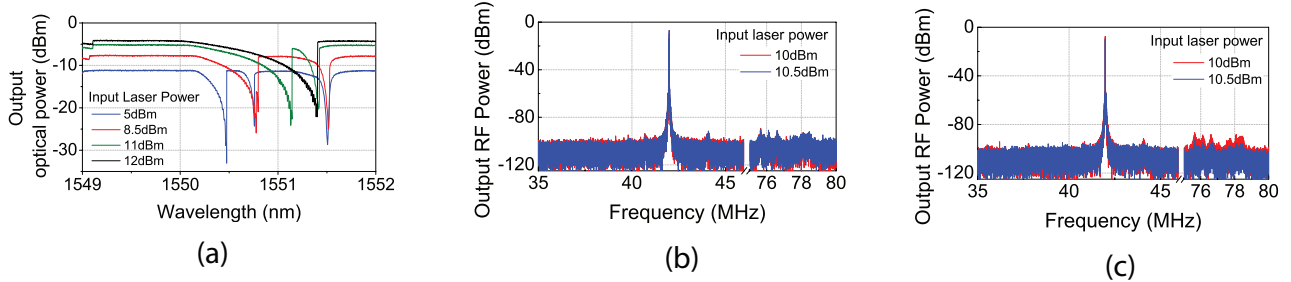}
\caption{(a) Optical spectrum of the silicon nitride opto-mechanical resonator for different input laser powers. The modal refractive indices of multiple optical mode families have different temperature dependence. The shark-fin shape of the optical resonances is attributed to thermal absorption. (b) When the laser light (1550.55nm) is blue detuned with respect to the pair of optical resonances around 1550.6nm, the RF power for the Brownian noise motion peaks when the laser power is raised from 10dBm (red curve) to 10.5dBm (blue curve). (c) RF spectrum when the laser (1550.587nm) is blue detuned with respect to one optical resonance and red detuned with respect to the other. In this case, the fundamental radial mode of vibration at 41.97MHz is heated as the pump laser power is increased from 10dBm (red curve) to 10.5dBm (blue curve), while a group of azimuthal composite mechanical modes centered around 77MHz is cooled.}
\label{heatcool}
\end{figure*}

\subsection{\label{sec:level2}Simultaneous heating and cooling of two mechanical modes}

Interaction of mechanical modes of nanomechanical resonators with multiple optical modes has been shown before \cite{kippnems}, which enables both heating and cooling. However, this scheme causes either heating or cooling of the mechanical mode, depending on which optical resonance is pumped. Here we explore the feasibility of using two closely spaced whispering gallery modes to simultaneously achieve heating of one mechanical mode while cooling another mechanical mode using a single pump laser. The micro-ring resonator heats up due to thermal absorption as the laser power is increased, which leads to the characteristic shark fin optical spectrum \cite{carmonstable} owing to temperature dependence of the modal refractive index. Due to the rich mode spectrum of a silicon nitride micro-ring resonator, situations may arise where the resonator has multiple optical mode families. The modal refractive indices of these mode families may have different temperature dependence. As shown in Figure \ref{heatcool}.a, one of the optical modes is far more sensitive to the laser power. As such, it is possible to fix the laser wavelength such that the pump laser light is red detuned with respect to one of the cavity modes and blue detuned with respect to the other in thermal equilibrium \cite{carmonstable}.

\begin{figure}[htbp]
\centering
\includegraphics[width = 8.5cm]{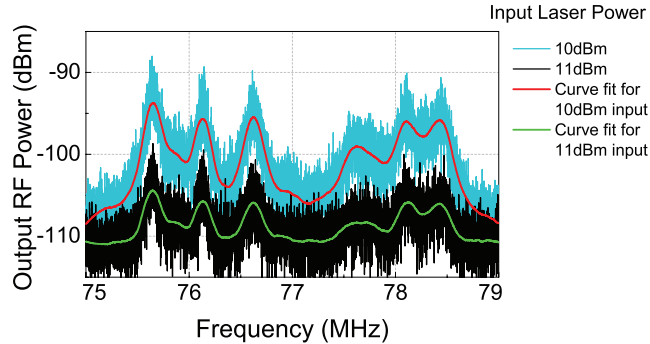}
\caption{Increasing the laser input power from 10dBm (blue curve) to 11dBm (black curve) results in cooling of the composite mechanical modes of the resonator. For instance, the linewidth of the mode at 76.7MHz increases from 150kHz to 250kHz by increasing the laser power. This corresponds to an effective temperature of 180K. The red and green curves are smoothed curve fits for the blue and black curves respectively.}
\label{coolzoom}
\end{figure}

Figure \ref{heatcool}.b shows increased RF power for the Brownian noise motion peaks when the laser power is increased. The laser light (1550.55nm) is blue detuned with respect to the pair of optical resonances at 1550.6nm. Figure \ref{heatcool}.c shows the RF spectrum when the laser (1550.587nm) is blue detuned with respect to one optical resonance and red detuned with respect to the other. In this case, the fundamental radial mode of vibration is heated as the pump laser power is increased while a group of azimuthal composite mechanical modes is cooled. Figure \ref{coolzoom} shows the cooling of these modes more clearly, with the linewidth for the mode at 76.7MHz increasing from 150kHz to 250kHz as the laser power is increased from 10dBm to 11dBm. This corresponds to an effective temperature of 180K.

\section{\label{sec:level1}Discussion}
The possibility of simultaneous heating and cooling of mechanical modes can open up avenues in studying coherent phonon exchange and phonon dynamics between different acoustic modes, mediated by an optical media and enable significant advances in ultra-precise sensing. Utilizing the damped mode for sensing in an all opto-mechanical transduced gyroscope promises high sensitivity and high resolution, while maintaining large bandwidth and high dynamic range. MEMS gyroscopes have attracted increasing attention from the automotive and defense industries in the past decade. The future expansion of their application domain relies on better resolution, sensitivity and stability. To improve the resolution and sensitivity, it is desired to use a high mechanical Q sense mode, as the response in the sense mode is amplified by the mode quality factor. Moreover, the fundamental noise floor is reduced, because the thermal noise associated with the mode is inversely proportional to the quality factor. The trade-off is reduced bandwidth, limited dynamic range and increased difficulties for matching the frequencies of the sense mode and the drive mode to cancel error sources. A close-loop design using electrostatic force feedback can mitigate the bandwidth and dynamic range problems. However, each stage of electronic circuitry adds additional noise to the system. Cooling the sense mode broadens the bandwidth of the mechanical resonator without adding extraneous noise. Cooling multiple closely spaced mechanical modes to groundstate will also provide an exciting toolset for studying aspects of condensed-matter and many-body physics at the macro-scale.

\begin{acknowledgments}
The authors wish to thank Suresh Sridaran for helpful discussions with the design and fabrication of the device, and David Hutchison for help with the experimental setup. This work was supported by the DARPA ORCHID program and Intel Academic Research Office. The devices were fabricated at the Cornell NanoScale Science and Technology Facility.

\end{acknowledgments}


\end{document}